\pdfoutput=1
\documentclass[11pt]{article}

\usepackage[preprint]{acl}

\usepackage{times}
\usepackage{latexsym}

\usepackage[T1]{fontenc}
\usepackage[utf8]{inputenc}

\usepackage{microtype}

\usepackage{inconsolata}

\usepackage{graphicx}

\usepackage{makecell}

\title{Implementation Considerations for Automated AI Grading of Student Work}

\author{
  \textbf{Zewei (Victor) Tian\textsuperscript{1}},
  \textbf{Alex Liu\textsuperscript{1}},
  \textbf{Lief Esbenshade\textsuperscript{1}},
  \textbf{Shawon Sarkar\textsuperscript{1}}
\\
  \textbf{Zachary Zhang\textsuperscript{2}},
  \textbf{Kevin He\textsuperscript{2}},
  \textbf{Min Sun\textsuperscript{1}}
\\
\\
  \textsuperscript{1}University of Washington,
  \textsuperscript{2}Hensun Innovation
}

\begin{document}
\maketitle
\begin{abstract}
This study explores the classroom implementation of an AI-powered grading platform in K–12 settings through a co-design pilot with 19 teachers. We combine platform usage logs, surveys, and qualitative interviews to examine how teachers use AI-generated rubrics and grading feedback. Findings reveal that while teachers valued the AI’s rapid narrative feedback for formative purposes, they distrusted automated scoring and emphasized the need for human oversight. Students welcomed fast, revision-oriented feedback but remained skeptical of AI-only grading. We discuss implications for the design of trustworthy, teacher-centered AI assessment tools that enhance feedback while preserving pedagogical agency.
% 19 K-12 teachers participated in a co-design pilot study of an AI education platform, testing assessment grading. Teachers valued AI’s rapid narrative feedback for formative assessment but distrusted automated scoring, preferring human oversight. Students appreciated immediate feedback but remained skeptical of AI-only grading, highlighting needs for trustworthy, teacher-centered AI tools.
\end{abstract}

\section{Introduction}

The integration of artificial intelligence (AI) into K-12 education has shown promise but also comes with new challenges \citep{wang_artificial_2024}. AI-powered educational platforms can offer tools to create instructional materials as well as to provide grading and feedback for assessments. Such tools purport to streamline workflows and provide rapid, individualized feedback. However, concerns arise regarding the alignment with pedagogical goals, the preservation of teacher agency, and mixed impacts on learners. This study engaged 19 teachers in a co-design pilot study for Colleague AI, an online AI-powered education platform for teachers and students that provides AI-based classroom functionality. In this study we focus on the AI grading and feedback functionality and provide generalizable information about how teachers envision the successful implementation of such a tool. By combining quantitative usage data with thematic analyses of teacher interviews and surveys, we examine the conditions under which AI-powered grading practices can augment instructional expertise of the educators. We situate our findings within the broader context of standard-based grading (SBG) and formative feedback theory, elaborating the opportunities AI tools can offer with actionable insights for developers, educators, school leaders and other stakeholders who are committed to empowering education through the assistance of AI without compromising instructional integrity.

\subsection{Historical Development of Automated Grading Systems}

Automated grading systems have a rich history spanning over nearly a century. In the 1940s, IBM introduced tabulating and test-scoring machines to accelerate scoring, reporting, and computing of assessments \citep{lorge_tabulating_1942}. This system was considered to be helpful in saving teachers time and processing student data more efficiently \citep{benham_method_1962}, and marks a significant early step toward automated grading. In the 1990s and entering into the 21st century, the introduction of learning management systems (LMS) brought automated grading into the spotlight, together with other functionalities around managing and distributing assessments. With education practices shifting to the digital realm, automated grading systems were driven to improve and adapt. Advanced technological innovations like natural language processing (NLP) and computer vision also assisted in the development of automated grading \citep{jocovic_automated_2024, ramesh_automated_2022}. However, as K-12 education adopts standard-based grading (SBG), assessments, especially formative assessments, require more complicated and comprehensive grading practices. Adding on to that, automated grading primarily focused on handling multiple choice questions while scoring open-ended questions like essays still remains a challenge \citep{ramesh_automated_2022}. In this context, the emergence of Large Language Model (LLM) Artificial Intelligence (AI) systems shows a potential next step in the integration of pedagogical frameworks and automated grading systems, and enables the automated grading process to be adopted on various assessment forms \citep{chu_llm-powered_2025, li_llm-based_2025, liew_automated_2025}. 

\subsection{The Value of Formative Assessment and Feedback in K-12 Education}

Effective assessment in K-12 education measures student learning while catalyzing continued growth. However, translating these principles into classroom practice faces practical challenges. This section examines theoretical foundations and empirical evidence supporting formative assessment in K-12 settings, while acknowledging systemic barriers that prevent educators from implementing these practices at scale.

\subsubsection{Rubric and standard-based grading in K-12}

K–12 education has increasingly adopted standards‐based grading (SBG) systems that align with state learning standards, shifting the focus from accumulating points to demonstrating mastery of specific competencies \citep{guskey_developing_2001, munoz_standards-based_2015}. By reporting student performance in terms of proficiency levels—such as “emerging,” “developing,” “proficient,” and “advanced”—SBG provides educators and families with a clearer picture of where learners stand relative to defined objectives \citep{oconnor_repair_2007}. SBG rubrics feature criteria appropriate to an assessment’s purpose and describe these criteria across a continuum of performance levels, ensuring that each standard is assessed with both clarity and precision \citep{brookhart_appropriate_2018}. When rubrics are crafted in alignment with state or district standards, they serve as the bridge between curricular goals and day‐to‐day classroom tasks \citep{mctighe_essential_2013}.
In K–12 settings, rubrics serve multiple purposes. First, they clarify expectations for students by defining what knowledge and skills constitute “proficient” and “exemplary” work; knowing these distinctions helps students set concrete targets and engage in self‐assessment \citep{andrade_teaching_2005, chowdhury_application_2018}. Second, rubrics provide consistent grading criteria for teachers, reducing subjectivity and inter‐rater variability. Rubric‐based scoring enhances reliability across different instructors and class sections \citep{jonsson_use_2007}. Finally, rubrics facilitate communication with parents about student progress: when teachers share rubric scores or performance descriptors, families gain concrete insight into their child’s strengths and areas for growth, enabling more focused conversations about how to support learning at home \citep{chowdhury_application_2018, popham_transformative_2011}.

\subsubsection{Timing and effectiveness for young learners}

Research shows that feedback timing critically affects K–12 learning \citep{ruiz-primo_examining_2013}. A meta-analysis reports: “feedback is one of the most powerful influences on learning and achievement” \citep{hattie_power_2007}.  Immediate feedback prevents misconceptions from becoming reinforced, which is particularly important for young learners building foundational skills. Students who receive immediate feedback during tasks retain information better and correct errors faster than those given delayed feedback \citep{ajogbeje_enhancing_2023}. These effects are evident across subjects like math and science, where rapid corrective guidance maintains motivation and supports mastery \citep{dihoff_provision_2004, mandouit_revisiting_2023}.
However, in many K–12 classrooms, practical constraints make providing immediate feedback difficult to sustain. Providing formative feedback to an entire class requires teachers to collect, analyze, and respond to each student’s work—a process that research shows is hard to implement at scale and sustain over time \citep{hopfenbeck_challenges_2023}. Moreover, a 2024 RAND survey of K–12 educators found that inconsistent access to formative-assessment tools—such as LMS-integrated grading, handheld response devices, or classroom response systems—forces many teachers to rely on paper-based workflows and delay feedback until weekly or biweekly grading cycles \citep{doan2024american}. A 2025 survey of 254 K-12 teachers found that although most value immediacy, workload and inconsistent access to digital tools prevent real-time feedback delivery. Without embedded systems (e.g., response-clickers or automated grading), teachers default to batch feedback, reducing impact \citep{jin_feedback_2025}. As a result, feedback often arrives days after submission, by which point students have moved on to new material, weakening the corrective value and allowing misconceptions to persist until the next evaluation cycle.

\subsubsection{Separating Formative Feedback from Evaluative Grades in K-12}

Separating formative feedback from evaluative grades is essential in K–12 education to prioritize learning and development over ranking. In a classic experimental study, the result showed that sixth-grade students who received detailed comments without grades demonstrated higher intrinsic motivation and better task performance than peers who received grades or grades paired with comments \citep{butler_effects_1986}. Another study found that when grades accompany comments, students tend to focus on the grade itself and disregard substantive feedback \citep{black_assessment_1998}. When feedback is decoupled from grades, teachers can devote attention to describing specific strengths, identifying misconceptions, and suggesting corrective steps without students fixating on scores \citep{brookhart_gathering_2022, wiliam_embedded_2011}. 

\subsubsection{Recent Research on Automated Graders and Real-Time Feedback in K-12}

Recent AI advancements have begun to extend assessment capabilities in K–12 contexts, but implementation in K-12 schools has typically lagged behind higher education. For example, M-Powering Teachers is an automated feedback tool that utilizes natural language processing to analyze verbal classroom interactions and subsequently provides formative feedback to teachers. In a randomized controlled trial with over 1,100 instructors in an online computer science course, the tool increased instructors’ use of “uptake” practices (i.e., acknowledging and building on student ideas) by 13 percent \citep{demszky_can_2024}. This result suggests promise for providing feedback to K-12 teachers to improve their classroom practices with AI-assisted analysis of their teaching
This also applies to other activities like administering assessments in the classroom. AI-assisted grading systems are being developed to analyze assessments and provide standard-based rubric \citep{tian_rubric_2025}, which then will be used to generate grades and feedback aligned with the standards. These tools recognize the unique needs of K-12 education, including age-appropriate feedback and alignment with Common Core and state standards.
However, limitations persist in K-12 contexts. Systematic scoping reviews note that AI tools often assume mature organizational structures and language conventions, which younger learners have not yet mastered \citep{lindsay_responsible_2023, yan_practical_2024}. Moreover, K–12 educators express concerns that AI‐mediated feedback may not sufficiently address younger students’ socio-emotional needs or align with grade-level curricula—barriers that slow adoption in elementary schools \citep{castro_implementation_2025, lin_engaging_2021}.

\section{Sample \& Methods}

For this study, we ran a seven week co-design pilot study with twenty-one teachers from four public school districts and one independent school in the Puget Sound region of Washington state to test the use of an AI powered learning platform’s student facing classroom features. Nineteen teachers participated in implementing and testing the assessment feature with their classrooms. Teachers participated in weekly discussion sessions where they received guidance about the platform, discussed how they might use the platform in their classrooms, and provided feedback about how they used the platform. Teachers completed weekly surveys about their platform usage. Two weeks of the pilot study focused on assessment grading. For this study we focus on the usage of the assessment grading functionality. In the pilot study, we interacted directly with teachers as they tested the platform in their classrooms. Students were not the subject of the study, and researchers did not directly interact with students. The study was approved by the University of Washington Institutional Review Board. 

During this phase of the study teachers were asked to implement two assessments in their classroom using the platform. Implementation of an assessment comprised several steps. First, teachers defined the purpose, type, and content of the assessment that they would give to their students. Teachers were instructed to only give assessments that fit with their classroom goals and that fit with their regular teaching practice. Then, teachers were given the option to use the AI platform to design a rubric to accompany their assessment and assist in providing feedback to their students. Once students completed the assessment, teachers had the option to allow students to view AI generated feedback and resubmit their assignment (i.e. a formative use) or to allow students only a single submission (i.e. a summative use). Finally, teachers reviewed the AI generated feedback - teachers were able to see the AI generated feedback whether or not they chose to allow students to view it - and returned their own feedback and grades to their students. 

\begin{figure*}[htbp]
  \includegraphics[width=\linewidth]{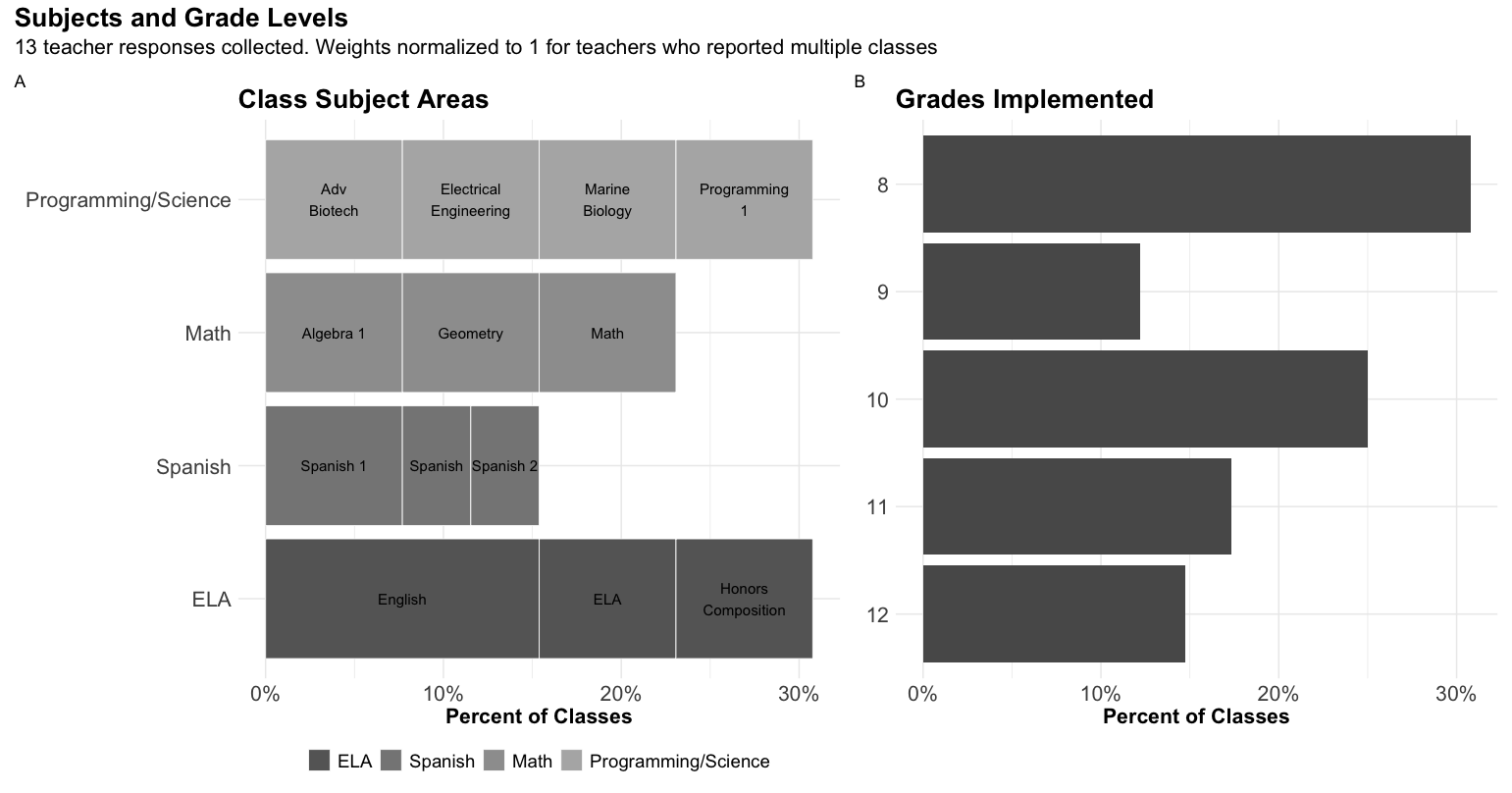}
  \caption{}
  \label{fig:subject_grade}
\end{figure*}

Teachers submitted surveys on how the implementation went. Of the 19 teachers who participated in the assessment tool portion of the study, 13 submitted feedback survey forms detailing how they used the platform to implement assessments. The implementations covered a range of class subject areas including programming/science courses (30\%), Math classes (25\%), Spanish language classes (15\%) and ELA classes (30\%). Classes were divided between grades 8 through 12. See Figure \ref{fig:subject_grade} for the full breakdown. Some teachers reported trying the tool in multiple class sections, because individual teachers are the focal unit of the study, we have weighted the responses such that each teacher counts equally (e.g. if teacher A reported a single math class and teacher B reported 2 English classes, we would report that study comprised half math and half English classes). 

\begin{figure*}[htbp]
  \includegraphics[width=\linewidth]{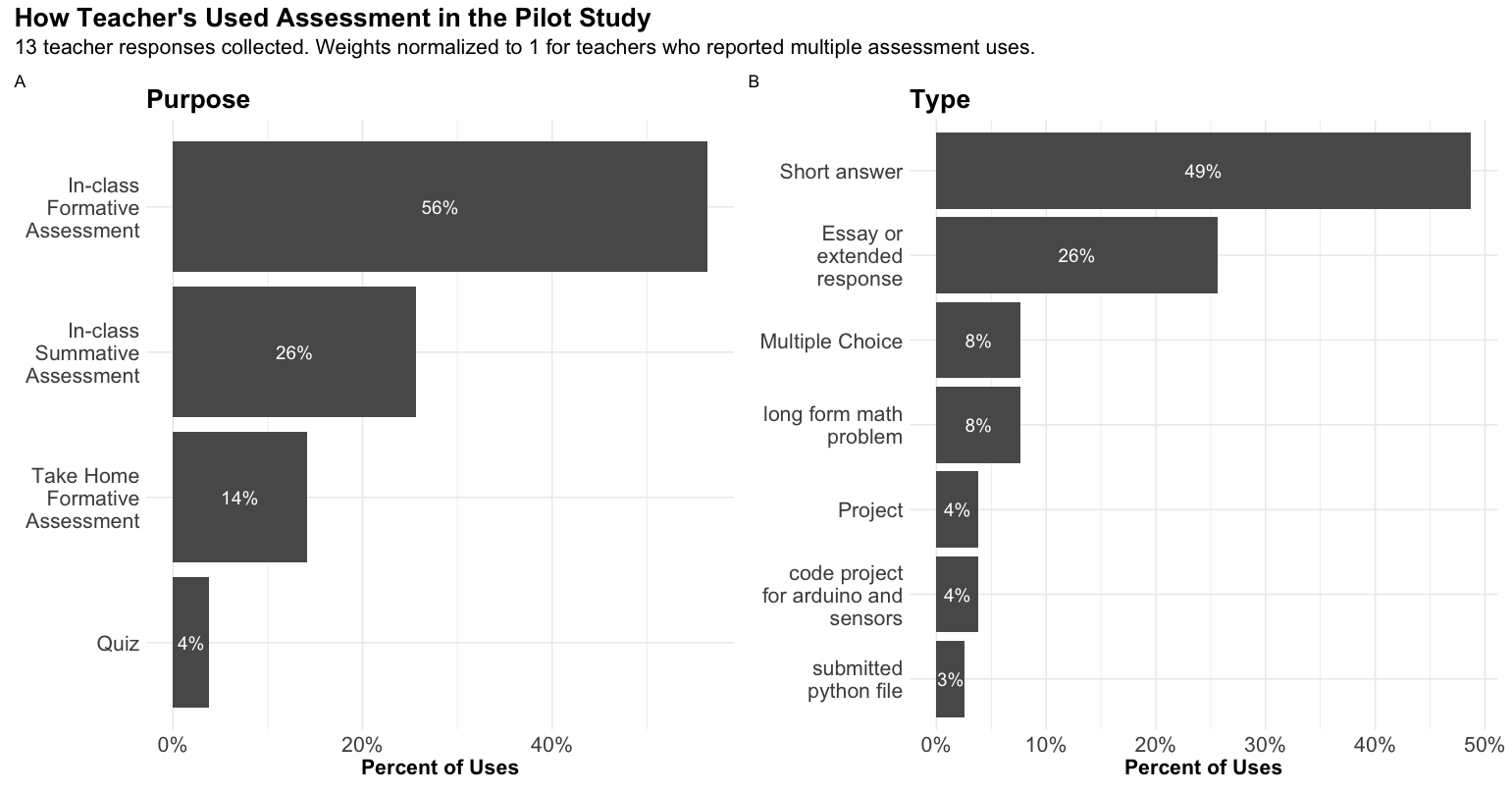}
  \caption{}
  \label{fig:purpose_type}
\end{figure*}

Figure \ref{fig:purpose_type} summarizes the type and purpose of the assessments given. Over half (56\%) of teachers who repsonded to the survey used the AI platform to administer an in-class formative assessment, and almost half (49\%) had ‘short-answer’ type questions in the assessment.  Although only 13 teachers completed the survey, 19 teachers did implement at least one assessment in their classrooms. In total, assessments were created in 33 unique classrooms with 936 student works submitted. 

In addition to requesting structured feedback in surveys on the implementation of the AI Grading tool, we applied thematic analysis to qualitative data sources including open-ended survey responses, group discussions, and individual interviews. We employed ground theory to thematic coding \citep{braun_using_2006} and identified recurring experiences, affordances, barriers, and recommendations from teachers’ perspectives. The established codebook (Appendix \ref{sec:codebook}) contains 7 parent code and 18 child code illustrating teachers’ and students’ user experiences from pilot teachers’ perspectives.

We also examined platform log data to understand the scope of the classroom implementation of the AI Grading tool, recording the number of assignments created, the number of student submissions made, whether students resubmitted their assignment and whether the teacher used the platform to return feedback to students. 

\section{Results}
\subsection{Platform Log Data Analysis}
The platform log dataset includes assessment logs from 33 unique classrooms created by 19 teachers. On average, each classroom implemented approximately 1.76 assessments. From the platform-generated assessment logs from 58 assessments, we conducted usage analysis to capture how AI grading and feedback features were implemented across subjects and school sites. The logs included information on total student enrollment in the classroom, number of submissions, AI-graded assessments, and resubmission counts.

\subsubsection{Submission Patterns and Engagement}
Submission rates varied widely, with a mean submission rate of 54.8\% (SD = 27.9\%). While some classrooms achieved full participation, others showed near 0 submission rates, indicating variability in how assessment activities were adopted across contexts. This variation reflects both instructional choice and logistical constraints (e.g., class type, student access, timing).

\subsubsection{AI Grading Coverage and Automation}
AI systems graded the majority of submitted assessments. In over 75\% of classrooms, more than 80\% of submitted student work received AI-generated scores. The median AI grading coverage was 92.2\%, with many classrooms achieving near-total automation. Both teachers and students can initiate AI grading to generate feedback and evaluation. This high rate of automated grading illustrates the system’s capacity to streamline evaluation workflows at scale.

\subsubsection{Student Resubmission Behavior}

Resubmissions, which may indicate iterative learning or clarification efforts, were relatively infrequent but nontrivial. On average, 8.7\% of students submitted work more than once, with a maximum observed rate of 66.7\% in one classroom. While not ubiquitous, this behavior suggests some teachers and students leveraged the platform’s capacity for revision and feedback loops.

\begin{table}[h!]
\centering
\begin{tabular}{|l|r|}
\hline
\textbf{Metric} & \textbf{Value} \\
\hline
Unique Classrooms & 33 \\
Average Assessments per Classroom & 1.76 \\
Mean Submission Rate & 54.8\% \\
\makecell[l]{Median AI Grading Coverage \\ on Submitted Works} & 92.2\% \\
Average Resubmission Rate & 8.7\% \\
\hline
\end{tabular}
\caption{Summary of Assessment Metrics. \newline
\textit{Note: Metrics are based on platform logs from 58 classroom-level assessment records across middle and high school implementations.}}
\end{table}

AI-powered grading was widely implemented across classrooms, with most student work receiving automated scores. Yet the variability in student engagement, along with uneven resubmission activity, reinforces a central finding from our qualitative analysis: teacher mediation remains essential to interpreting and contextualizing AI output. Teachers did not simply deploy automation, instead they integrated it into their classroom practices to balance speed with pedagogical intent.

\subsection{Teacher Survey Data}

13 out of 19 teachers returned survey forms about their use of the AI Grading platform. In addition to the data on the implementation context, they also evaluated the quality of the AI generated rubrics and the AI generated feedback. 

\subsubsection{Rubric Quality}

Over 60\% of the teachers indicated that they were able to use the AI generated rubrics in their classroom assignments. The majority indicated that they made minor changes to the rubric, indicating that they were not willing to fully accept the AI generated content without review and adjustment. Interestingly, no teachers indicated that they made major changes to the AI generated rubric. Only 7\% of teachers indicated that the rubrics could not be used in their classroom - either because they needed major revisions or were simply not applicable. Roughly a quarter of teachers reported that they did not attempt to use the AI generated rubrics at all. Note that some teachers submitted multiple response forms for their different classrooms, the overall results are weighted so that each teacher has equal weight. 

\begin{figure*}[t]
  \includegraphics[width=\linewidth]{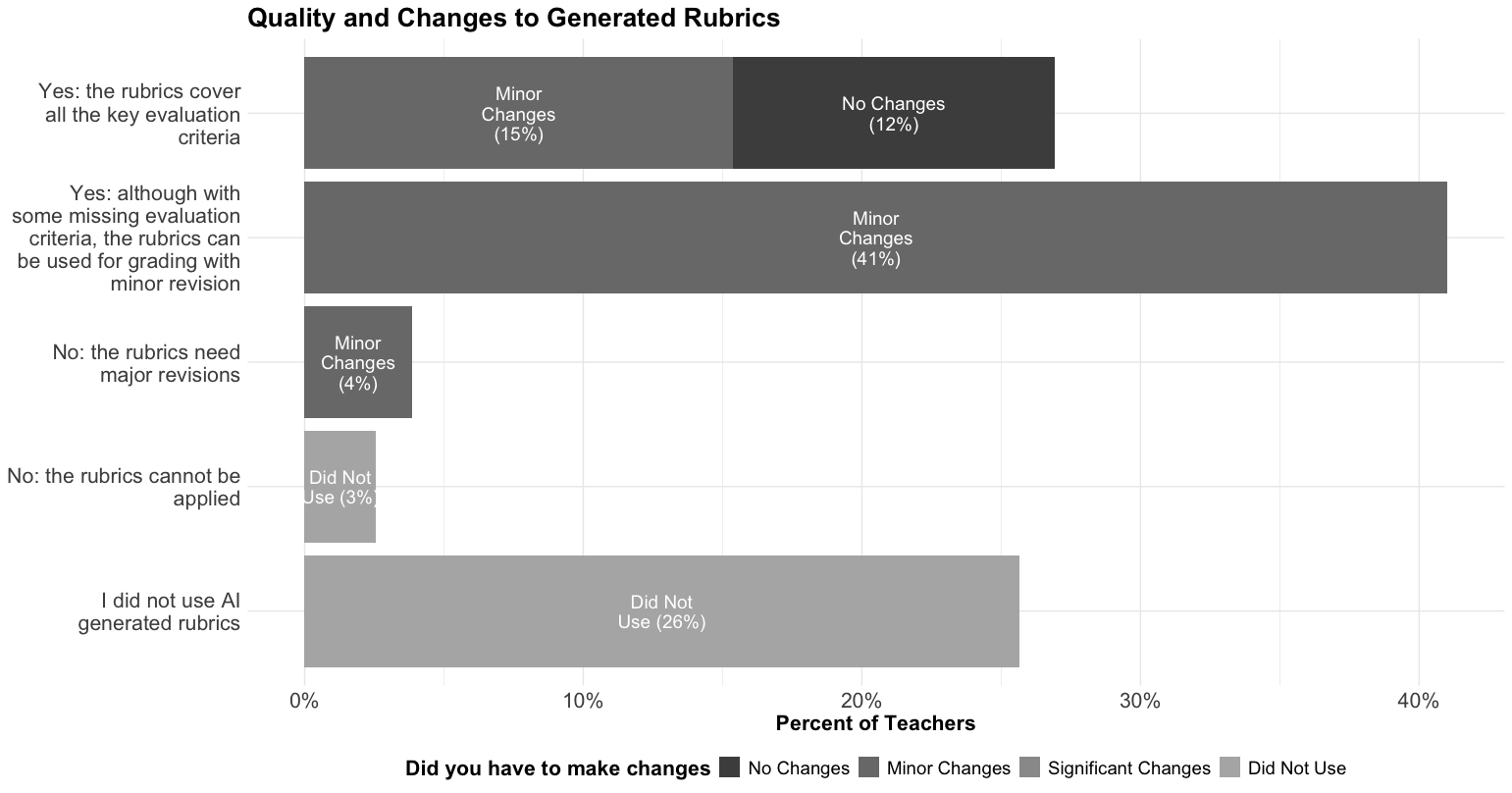}
  \caption{Teacher survey responses to the quality of the AI generated rubric and whether it was necessary to make changes. }
  \label{fig:rubric_quality}
\end{figure*}

\subsubsection{AI Generated Feedback Quality}

57\% of teachers indicated that the AI feedback provided clear, actionable feedback for teachers or students, with 41\% indicating that the feedback was useful for both teachers and students, 14\% indicating that the feedback was only useful for students, and 3\% indicating that it was only useful for teachers.  42\% of teachers indicated that the feedback was not useful, with 24\% indicating that the feedback was vague or unhelpful and 18\% indicating that it was incorrect or misleading. 

\begin{figure*}[t]
  \includegraphics[width=\linewidth]{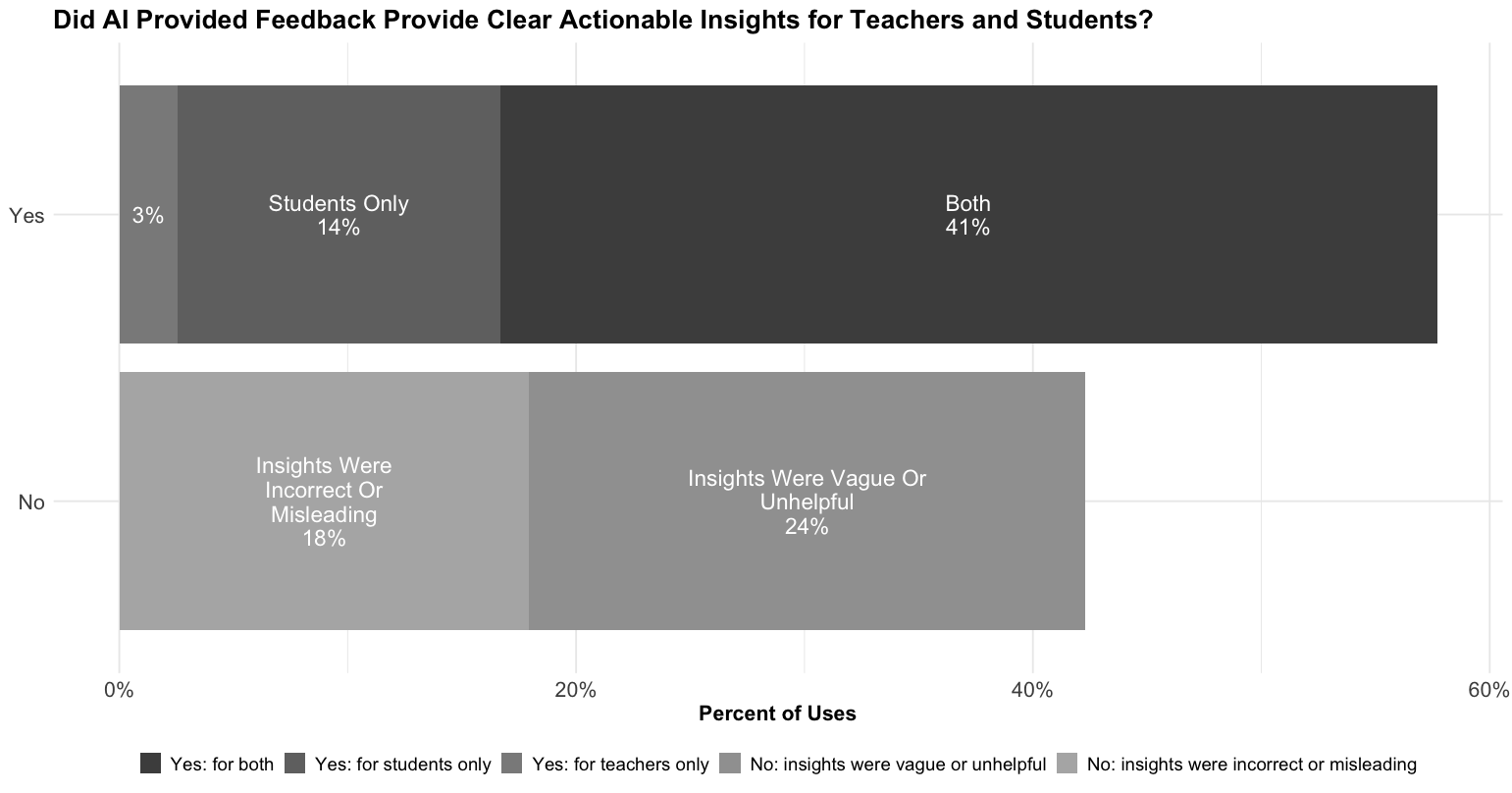}
  \caption{Teacher survey responses to the quality of the AI generated assessment feedback}
  \label{fig:feedback_quality}
\end{figure*}

\subsection{Discussion Transcript \& Interview Qualitative Analysis}

To deepen our understanding of how teachers experienced the AI-powered assessment, grading, and feedback features in real K-12 classroom contexts, we conducted a qualitative analysis as well. The qualitative analysis yielded three central themes, reflecting both the promise of AI to enhance feedback workflows and the structural and pedagogical tensions that emerge in educational contexts. 

\subsubsection{AI Grading: Feedback as Formative Scaffold over Numerical Scores}

Across classroom contexts, teachers consistently emphasized the pedagogical value of narrative feedback over numerical grades. While the platform offered a mechanism for scoring, many teachers found the AI’s application of point values to be inconsistent or misaligned with their rubrics. One teacher shared, “The tool scored some students out of 20 points and others out of 10, when I had specified the assessment was worth 10 points” (Marine Biology, Grades 10–12). Others noted the AI “took points off for things not in the rubric” or used standards “outside of the students’ current skill level” (Engineering, Grades 9–12).
By contrast, the system’s narrative feedback was frequently praised for its specificity, clarity, and alignment with formative goals. Teachers described it as a useful “first draft” that helped identify student misconceptions and suggest improvement strategies. “While I found the feedback from the AI to be fairly accurate, it seemed inconsistent in terms of how it attached numbers to that feedback,” stated by the same grades 10–12 marine biology teacher.
This tension between qualitative and quantitative outputs suggests that current LLM-based assessment systems may be best positioned as formative tools, generating scalable, revisable feedback that scaffolds learning, rather than reliable summative graders. Teachers expressed interest in treating AI grading as a fast first-pass diagnostic, followed by human adjustment. “[Students] loved the prospect of getting a grade and feedback with such a quick turnaround, rather than waiting the 2–3 weeks that it usually takes me to grade their writing” (English, Grade 11).

This orientation toward feedback-first design reinforces the importance of transparency and explainability in AI-powered assessment tools. When numerical scores lack clarity or consistency, but written comments hold pedagogical value, the role of the AI should be reimagined: not to replace teacher judgment, but to scaffold learning through accessible, timely, and editable feedback.

\subsubsection{Teacher Oversight Enables Trust and Personalization}

Teachers reported that AI feedback, while efficient, was not passively accepted by students. Students actively scrutinized AI generated evaluations’ fairness, clarity, and alignment with their work. This dynamic created new expectations for teachers to engage in the grading and feedback process, not just as overseers, but as collaborators who could validate, revise, or clarify the AI’s output. One educator noted, “My students were confused why some feedback was so positive, yet the score was low. They came to me asking if the grade was accurate and what it really meant” (English, Grade 11). 
Far from seeing this as a burden, many teachers described this supervising and collaborative role as essential and empowering. It allowed them to reinforce instructional goals, personalize communication with students, and restore fairness to the grading process. “I took the feedback and put it in the AI Chat… told it, ‘give me one paragraph in teacher voice,’” one teacher explained, reflecting the effort to mediate AI output in a way that aligned with their teaching persona and classroom discourse (ELA, Grade 10). Another explained, “I asked if [students] would be ok if the AI graded all their work and they all said no! They want to know I’m reading their work. They want me to see their jokes and emotions. They feared that AI would just be like a checklist. I thought her [the AI assistants] feedback was better than mine though. (But they thought it’d be great as a pre-submission self check grade)” (Science, Grade 11). These moments highlight that personalization is not merely the product of generative automation, it is co-produced through educator framing and student trust.
Even among those critical of the AI’s limitations, teachers valued the system’s ability to streamline initial feedback, reduce turnaround time, and make space for higher-order instructional moves. “I see AI grading tools as a kind of new TA: it gives fast, helpful first-pass feedback that enables students to make improvements right away, but I still review and make final grading decisions.” (Engineering, Grade 9-12). Even teachers who were critical of the AI’s limitations noted its utility for surfacing initial insights that they could refine or expand. In this human-AI collaborative process, automation enhances efficiency, but teacher oversight ensures that outputs align with pedagogical goals and relational norms. Teachers stressed that speed alone was insufficient: the AI’s utility depended on whether its feedback meaningfully reflected classroom expectations. “The time-saving is great. But only if the comments represent how I would actually respond to student work. Otherwise I have to re-do it anyway.” (Math, Grades 7–8).
This convergence of student demand and teacher professional judgment highlights a collaborative model of assessment: one where AI tools extend instructional reach, but teachers retain interpretive authority. Personalization, in this view, is not the result of automation alone, it is made meaningful when filtered through pedagogical expertise and enacted in response to learner needs.

\subsubsection{Student Engagement is Mediated by Interface Design and Accessibility Considerations}

Teacher noted that student responses to AI evaluations varied significantly, shaped not only by content quality but also by interface design and prior technology exposure. On one hand, Several teachers reported strong engagement among struggling or anxious learners, who appreciated the opportunity to receive feedback before submitting to peers. For example, a grade 11 IB English teacher shares, “some of my struggling students… liked having someone to give feedback before sharing. It made them more confident.” On the other hand, some students were overwhelmed by the volume or complexity of the comments. One teacher noted, “They thought it was a lot of feedback... It might have been better to let me limit it to just a few things” (World Language, Grades 9–11). Technical usability also posed barriers: some students had trouble uploading assignments, locating relevant sections of the AI-generated feedback, or were put off by first impressions of the interface, which “looked old,” all of which may have led them to abandon the tool after initial attempts. 
The mixed reception reinforces the importance of usability design and accessibility. Without scaffolds for clarity and navigation, AI systems may inadvertently heighten disparities in experience and learning outcomes among students with different levels of digital fluency. To avoid these pitfalls, developers must prioritize transparency, explanation, and accessibility in system design. Features such as adjustable feedback volume, simpler and fashionable interfaces, and teacher-led onboarding may be critical to ensuring that AI systems support meaningful engagement across all learners.

\section{Discussion \& Conclusions}

This study offers early empirical insight into how K–12 educators engage with AI-powered grading systems in real classroom contexts. Through a co-design pilot with 20 teachers, we observed that while automated scoring tools are increasingly capable of streamlining feedback and assessment workflows, their successful classroom implementation hinges on how well they align with formative goals, support teacher expertise, and align with student expectations for fairness. Teachers used the platform to generate rubrics, assign assessments, deliver formative feedback, and manage revision cycles, but they did not treat AI output as final. Instead, they exercised discretion, editing feedback, clarifying grades, and recontextualizing comments to maintain pedagogical coherence. This model where automation accelerates routine processes but teachers retain interpretive control emerged as a key condition for productive use.

Throughout the study, three themes emerged: (1) teachers emphasized narrative feedback over numeric scores, valuing elaborated comments generated by AI that revealed misconceptions and next steps for learning; (2) teacher mediation was essential to address discrepancies between comments and grades, underscoring that AI should augment rather than replace educator judgment; and (3) student responses varied—some benefited from low-stakes feedback with a quick turnaround, while others experienced cognitive overload or usability challenges, revealing heterogeneity considerations tied to digital literacy and AI competency in modern classrooms.

While this pilot study provides valuable insight into teacher experiences with AI-powered grading tools, several limitations warrant consideration. First, the sample was geographically limited to the Puget Sound region and comprised volunteers who may be more open to AI technology use than the broader teaching population, potentially introducing selection bias. Second, not all participating teachers completed post-implementation surveys, which may skew those findings toward those with stronger opinions or more successful experiences. Third, the study relied on teacher self-reported data and platform logs rather than direct observation of classroom implementation, limiting our ability to assess actual student interaction with the AI system. Finally, this study is of a single generative AI based platform, and findings may not fully generalize to other AI grading and feedback systems. 

Despite its exploratory scope, this study yields several insights that are likely to generalize beyond the immediate implementation context. Most notably, teachers consistently valued AI-generated narrative feedback as a formative tool, even when they questioned the reliability of automated scoring. This suggests that LLM-based grading systems may be best positioned not as replacements for teacher judgment, but as scaffolds for feedback-rich instruction. Additionally, the finding that students desired teacher involvement—even when AI feedback was accurate—underscores the importance of maintaining human connection and interpretive authority in automated systems. Finally, the study highlights design considerations for future AI tools: systems should allow for teacher oversight, offer clear interfaces for student understanding, and support workflows that enable iterative revision. These features are likely to be essential across a wide range of school settings and instructional models.

\section*{Acknowledgments}
This work is supported by the Institute of Education Sciences of the U.S. Department of Education, through Grant R305C240012 and by several awards from the National Science Foundation (NSF \#2043613, 2300291, 2405110) to the University of Washington, and a NSF SBIR/STTR award to Hensun Innovation LLC (\#2423365). The opinions expressed are those of the authors and do not represent views of the funders.
\bibliography{latex/AIME}

\newpage
\clearpage

\onecolumn
\appendix

\section{Codebook of Educator Feedback on AI-Powered Assessment}
\label{sec:codebook}

\begin{table}[htbp]
\centering
\caption{Qualitative Coding Scheme with Frequencies}
\label{tab:comprehensive_coding}
\footnotesize
\begin{tabular}{|p{0.8cm}|p{2.8cm}|p{2cm}|p{1.2cm}|p{6cm}|}
\hline
\textbf{Code} & \textbf{Parent Code} & \textbf{Child Code} & \textbf{Frequency} & \textbf{Child Code Description} \\
\hline
WF1 & Workflow and Implementation & Feature Setup Challenges & 29 & Describes obstacles in setting up assessments, rubrics, or assignments using the AI tools. \\
\hline
WF2 & Workflow and Implementation & Alternative Use Cases & 8 & When teachers adapted or repurposed features for different pedagogical intents. \\
\hline
WF3 & Workflow and Implementation & Time-Saving Potential & 9 & Mentions of AI helping reduce grading load or turnaround time. \\
\hline
FB1 & Feedback Quality and Utility & Feedback Customization & 5 & Teachers modifying AI-generated feedback to suit student needs or tone. \\
\hline
FB2 & Feedback Quality and Utility & Feedback Usefulness & 12 & Teachers' perceptions of whether the feedback is pedagogically meaningful or accurate. \\
\hline
FB3 & Feedback Quality and Utility & Student Perception of Feedback & 2 & How students perceive or react to AI feedback. \\
\hline
ST1 & Student Impact & Increased Engagement & 5 & Positive changes in student engagement or willingness to revise based on AI feedback. \\
\hline
ST2 & Student Impact & Student Confusion or Frustration & 5 & Instances of student difficulty with interface, grading accuracy, or expectations. \\
\hline
ST3 & Student Impact & Equity of Support & 2 & Reflections on how AI tools affected different learner groups. \\
\hline
TR1 & Trust and Accuracy & Inconsistency of Grading & 2 & Reports of AI producing different results for the same submission or not aligning with rubric. \\
\hline
TR2 & Trust and Accuracy & Human Oversight & 10 & Emphasis on teacher's role in verifying or revising AI grading before finalizing. \\
\hline
UI1 & Usability and Interface & Clunky Interface or Poor UX & 2 & Descriptions of confusion or dissatisfaction with platform usability. \\
\hline
UI2 & Usability and Interface & Preferred Interaction Pathways & 2 & Teacher workarounds or preferences for using other tools. \\
\hline
PR1 & Professional Use and Reflection & Teacher Accountability and Editing & 1 & Teachers feeling responsible for editing and verifying AI output. \\
\hline
PR2 & Professional Use and Reflection & Planning for Growth & 4 & Teachers thinking about scaling or adjusting practice using AI. \\
\hline
SD1 & Suggestions for Development & Workflow Simplification & 2 & Recommendations to reduce clicks or streamline setup. \\
\hline
SD2 & Suggestions for Development & Granular Feedback Requests & 2 & Suggestions for item-level feedback or clearer linkage to rubrics. \\
\hline
SD3 & Suggestions for Development & Feature Expansion & 2 & Ideas like nudging systems, PDF exports, or data summaries by student. \\
\hline
\end{tabular}
\end{table}

\end{document}